\documentclass[11pt,a4paper]{article}

\usepackage{epsfig,amsmath,amssymb}

\begin{document}

\title{Thin-shell wormholes associated with global cosmic strings} 
\author{Cecilia Bejarano$^{1,}$\thanks{e-mail: cbejarano@iafe.uba.ar}, Ernesto F. Eiroa$^{1,}$\thanks{e-mail: eiroa@iafe.uba.ar}, 
Claudio Simeone$^{2,}$\thanks{e-mail: csimeone@df.uba.ar}\\
{\small $^1$ Instituto de Astronom\'{\i}a y F\'{\i}sica del Espacio, C.C. 67, 
Suc. 28, 1428, Buenos Aires, Argentina}\\
{\small $^2$ Departamento de F\'{\i}sica, Facultad de Ciencias Exactas y 
Naturales,} \\ 
{\small Universidad de Buenos Aires, Ciudad Universitaria Pab. I, 1428, 
Buenos Aires, Argentina}} 

\maketitle

\begin{abstract}
In this article we construct cylindrical thin-shell wormholes in the context of global cosmic strings. We study the stability of static configurations under perturbations preserving the symmetry and we  find that the throat tends to collapse or expand, depending only on the direction of the velocity perturbation.\\ 

\noindent 
PACS number(s): 04.20.Gz, 11.27.+d, 04.40.Nr

\end{abstract}

\section{Introduction}

The study of traversable Lorentzian wormholes has received great attention since the leading paper by  Morris and Thorne \cite{motho}. These objects are solutions of the equations of gravitation that consist in  two regions connected by a throat, which for static wormholes is a minimal area surface satisfying a flare-out condition \cite{ hovis1}. Wormholes can join two parts of the same universe or two separate universes \cite{motho, visser}, and they  must be threaded by exotic matter that violates the null energy condition \cite{motho, hovis1, visser, hovis2}. It was shown by Visser \textit{et al.} \cite{viskardad} that the amount of exotic matter needed around the throat can be made as small as desired by appropriately choosing  the geometry of the wormhole. A physically interesting class of wormholes is that of thin--shell ones, which are constructed by cutting and pasting two manifolds \cite{visser, mvis} to form a geodesically complete new one with a throat placed in the joining shell. In this case, the exotic matter needed for the existence of the configuration is located at the shell. Poisson and Visser \cite{poisson} performed a linearized stability analysis under spherically symmetric perturbations of a thin-shell wormhole made by joining two Schwarzschild geometries. Later, Barcel\'{o} and Visser \cite{barcelo} applied this method to  wormholes constructed using branes with negative tensions and Ishak and Lake \cite{ishak} analyzed the stability of transparent spherically symmetric thin-shells and wormholes. Eiroa and Romero \cite{eirom} extended the linearized stability analysis to Reisner-Nordstr\"{o}m thin-shell geometries, and Lobo and Crawford \cite{lobo} to wormholes with a cosmological constant. Also, Lobo and Crawford considered the stability of dynamical thin-shell wormholes \cite{lobo2}. Five-dimensional thin-shell wormholes in Einstein--Maxwell theory with a Gauss--Bonnet term were studied by Thibeault \textit{et al.} \cite{marc}. \\

Recently, there has been a renewed interest in the study of cosmic strings as a consequence of new developments in superstring theory which suggest that fundamental strings can play the role of cosmic strings, and also because of new observational evidence which may support their existence \cite{vi05}. In the presence of a complex scalar field, spontaneous symmetry breaking results in the formation of cosmic strings; these can be global or local, depending on the character of the symmetry which is broken in the phase transition \cite{vilbook}. In the case of local strings, a gauge field exists besides the complex scalar field, and the stress-energy tensor turns to be concentrated in a very thin cylinder. The gravitational effects of such objects have been extensively studied \cite{vilgohis}, its main features are that a constant deficit angle is induced in the space around, and the trajectories of non relativistic particles are not affected (in the absence of wiggles or kinks along the string \cite{vaas} and if a second phase transition leading to the appearance of charge carriers has not taken place \cite{wibapi}). The energy of global strings, instead, extends beyond the core, and static solutions with boost invariance along the symmetry axis have a physical singularity  at a finite distance \cite{coka,greg}.  If the condition of boost invariance is relaxed, the outer singularity can be avoided \cite{senba}. Global cosmic strings have near the core a deficit angle which grows with the radial coordinate, then decreases and finally turns into an excess angle that diverges at the singularity \cite{vilbook}. Also, the geodesics around a global string show the peculiarity of a repulsive effect \cite{harari} pushing matter towards the singularity \cite{coka}. These important differences make clearly non trivial an extension  to global strings of any analysis regarding the physics of local strings; in particular, wormholes associated with them deserve a separate and detailed study. \\

Wormholes associated with cosmic strings have been previously investigated by several authors. Cl\'{e}ment \cite{clem1} studied traversable multi-wormhole solutions where the metric asymptotically tends to the conical cosmic string metric, and later, he  extended cylindrical multi-cosmic strings metrics to wormhole spacetimes with only one region at spatial infinite, and he thoroughly analyzed  the geometry of asymptotically flat wormhole spacetimes produced by one or two cosmic strings \cite{clem2}. Aros and Zamorano \cite{arza} obtained a solution which can be understood as a traversable cylindrical wormhole inside the core of a global cosmic string.  Eiroa  and Simeone \cite{eisi} studied cylindrical thin-shell wormholes associated with local cosmic strings. In this article we analyze  thin-shell wormholes constructed from the spacetime metrics of global cosmic strings.  We do not intend to explain the mechanisms that might provide the exotic matter to them, but, instead, we focus on the geometrical aspects of these objects. In Sec. \ref{cylind} we construct the wormholes by applying the cut and paste formalism,  in Sec. \ref{stability}  we study the stability of the configurations under radial perturbations, and finally, in Sec. \ref{discu}, the results are discussed. Units where $c=\hbar=G=1$ are used.

\section{Thin-shell wormholes}\label{cylind}

We consider global strings with no currents present along the core and boost invariance along the string axis \cite{coka}. The metric of these cosmic strings, in coordinates $X^{\alpha}=(t,x,\theta ,z)$, takes the form  
\begin{equation}
ds^2 = f(x)(-dt^2 +dz^2 )+g(x) dx^2+ h(x) d\theta^2,
\label{e1}
\end{equation}
where 
\begin{equation}
f(x)=1-\frac{\ln x}{\ln x_{s}},
\label{e1a}
\end{equation}
\begin{equation}
g(x)=\gamma^2 e^{-\ln^2 x/\ln x_{s} }\left(1-\frac{\ln x}{\ln x_{s}}\right)^{-1/2},
\label{e1b}
\end{equation}
and
\begin{equation}
h(x)=g(x)x^2.
\label{e1c}
\end{equation}
This metric is valid outside the core, where the stress-energy tensor due to a Goldstone boson field has the form $T_{\mu\nu}=\partial_{\mu}\phi^{*}\partial_{\nu}\phi-(1/2)g_{\mu\nu}|\partial\phi|^2+h.c.$, with $\phi=Fe^{i\theta}$. The parameter $\gamma$ can be determined by matching the metric given above with the metric inside the core. The radius of the core is $x_{core}\approx 1$ and an outer physical singularity \cite{coka} occurs  at $\ln x_{s}=(8\pi F^2)^{-1}$, with $F$ determined by the scale of the symmetry breaking leading to the appearance of the topological defect. For $F$ much less than the Planck mass $M_P$, $x_{s}$ is very large; for example, when $F \approx 2\times 10^{17}GeV$, then $\ln x_{s}=137$ and  the singularity is situated at a distance of the order of the cosmological horizon \cite{coka}. The area per unit of $z$ coordinate for a fixed value of  $x$ is given by $4\pi\sqrt{f(x)h(x)}$. This area is an increasing function of $x$ from $x_{core}$ to $x_{wh}=x_s\exp (-\sqrt{\ln x_s}/2)$ and decreases with $x$ from $x_{wh}$ to $x_s$, where it is zero and the outer singularity is reached. Starting from the geometry given by Eq. (\ref{e1}),  we choose a radius $a$ between $x_{core}$ and $x_{wh}$, and we take two copies  of the region with $x \geq a$:
\begin{equation}
  \mathcal{M}^{\pm} = \{ X / x \geq a \}, \label{e2}
\end{equation}
and paste them at the hypersurface defined by
\begin{equation}
  \Sigma \equiv \Sigma^{\pm} = \{ X / x - a = 0 \}, \label{e3}
\end{equation}
to make a geodesically complete manifold $\mathcal{M}=\mathcal{M}^{+} \cup \mathcal{M}^{-}$. The area per unit of $z$ coordinate defined above is an increasing function for $x\in [a,x_{wh})\subset(x_{core},x_{wh})$; therefore the flare-out condition is satisfied and this construction creates a cylindrically symmetric thin-shell wormhole with two regions connected by a throat at $\Sigma $\footnote{If, besides, one demands that geodesics within a plane orthogonal to the string open up  at the throat, then the condition that $h(x)$ is an increasing function of $x$ is also required.}. On the manifold $\mathcal{M} $ we can define a new radial coordinate $l = \pm \int_a^x \sqrt{g(x)} dx$, where the positive and negative signs correspond, respectively, to $\mathcal{M}^+$ and $\mathcal{M}^-$, with $|l|$ representing the proper radial distance to the throat, which is placed in $l = 0$. To study this traversable wormhole we use the standard Darmois-Israel formalism \cite{daris,mus}. The throat of the wormhole is placed at the shell $\Sigma$, which is a synchronous timelike hypersurface. We can adopt coordinates $\xi ^i=(\tau,\theta,z )$ in $\Sigma $, with $\tau $ the proper time on the shell. For the analysis of stability under perturbations preserving the symmetry, we let the radius of the throat be a function of the proper time, that is $a = a ( \tau )$. The position of the throat is then given by the equation 
\begin{equation}
  \Sigma : \mathcal{H} (x, \tau ) = x - a ( \tau ) = 0. \label{e5}
\end{equation}
The extrinsic curvature associated with the two sides of the shell is:
\begin{equation}
  K_{ij}^{\pm} = - n_{\gamma}^{\pm} \left. \left( \frac{\partial^2
  X^{\gamma}}{\partial \xi^i \partial \xi^j} + \Gamma_{\alpha \beta}^{\gamma}
  \frac{\partial X^{\alpha}}{\partial \xi^i} \frac{\partial
  X^{\beta}}{\partial \xi^j} \right) \right|_{\Sigma}, \label{e6}
\end{equation}
where $n_{\gamma}^{\pm}$ are the unit normals ($n^{\gamma} n_{\gamma} = 1$) to
$\Sigma$ in $\mathcal{M}$:
\begin{equation}
  n_{\gamma}^{\pm} = \pm \left| g^{\alpha \beta} \frac{\partial
  \mathcal{H}}{\partial X^{\alpha}} \frac{\partial \mathcal{H}}{\partial
  X^{\beta}} \right|^{- 1 / 2} \frac{\partial \mathcal{H}}{\partial
  X^{\gamma}} . \label{e7}
\end{equation}
In the orthonormal basis  $\{ e_{\hat{\tau}}, e_{\hat{\theta}}, e_{\hat{z}}\}$, defined by $e_{\hat{\tau}} = \sqrt{1/f(x)}e_{t}$, $e_{\hat{\theta}} 
= \sqrt{1/h(x)}e_{\theta}$, $e_{\hat{z}} =\sqrt{1/f(x)}e_{z}$, the metric has the form
$g_{\hat{\imath} \hat{\jmath}} = \eta_{\hat{\imath}\hat{\jmath}}=diag(-1,1,1)$; thus the non vanishing components of the extrinsic curvature read
\begin{equation}
K_{\hat{\tau}\hat{\tau}}^{\pm}= \mp\frac{\sqrt{g(a)}}{2\sqrt{g(a)\dot{a}^{2}+1}}\left\{2\ddot{a} +\dot{a}^2\left[\frac{f'(a)}{f(a)}+\frac{g'(a)}{g(a)}\right] +\frac{f'(a)}{f(a)g(a)} \right\},
\label{e8a}
\end{equation}
\begin{equation}
K_{\hat{\theta} \hat{\theta}}^{\pm} =\pm\frac{h'(a)\sqrt{g(a)\dot{a}^{2}+1}}{2h(a)\sqrt{g(a)}} , 
\label{e8b}
\end{equation}
and
\begin{equation}
K_{\hat{z} \hat{z}}^{\pm} =\pm\frac{f'(a)\sqrt{g(a)\dot{a}^{2}+1}}{2f(a)\sqrt{g(a)}} , 
\label{e8c}
\end{equation}
where the dot stands for  the derivative with respect to $\tau$.
The Einstein equations on the shell (Lanczos equations) take the form:
\begin{equation}
-[K_{\hat{\imath} \hat{\jmath}}]+[K]g_{\hat{\imath} \hat{\jmath}}=8\pi S_{\hat{\imath} \hat{\jmath}},
\label{e10}
\end{equation}
where $[K_{_{\hat{\imath} \hat{\jmath}}}]\equiv K_{_{\hat{\imath}
\hat{\jmath}}}^+ - K_{_{\hat{\imath} \hat{\jmath}}}^-$, 
$[K]=g^{\hat{\imath} \hat{\jmath}}[K_{\hat{\imath} \hat{\jmath}}]$ is the 
trace of $[K_{\hat{\imath} \hat{\jmath}}]$ and
$S_{_{\hat{\imath} \hat{\jmath}}} = \text{\textrm{diag}} ( \sigma, 
-\vartheta_{\theta }, -\vartheta_{z} )$ is the surface stress-energy tensor, 
with $\sigma$ the surface energy density and $\vartheta_{\theta , z}$ the 
surface tensions. Then replacing Eqs. (\ref{e8a}), (\ref{e8b}) 
and (\ref{e8c}) in Eq. (\ref{e10}) we have
\begin{equation}
\sigma =-\frac{\sqrt{g(a)\dot{a}^{2}+1}}{8\pi\sqrt{g(a)}}\left[\frac{f'(a)}{f(a)}
+\frac{h'(a)}{h(a)}\right],
\label{e11}
\end{equation}
\begin{equation}
\vartheta_{z} = -\frac{\sqrt{g(a)}}{8\pi\sqrt{g(a)\dot{a}^{2}+1}} \left\{2\ddot{a}  +\dot{a}^2\left[\frac{f'(a)}{f(a)}+\frac{g'(a)}{g(a)}+\frac{h'(a)}{h(a)}\right]
+\frac{1}{g(a)}\left[\frac{f'(a)}{f(a)}+\frac{h'(a)}{h(a)}\right]\right\},
\label{e12}
\end{equation}
\begin{equation}
\vartheta_{\theta}  = -\frac{\sqrt{g(a)}}{8\pi\sqrt{g(a)\dot{a}^{2}+1}}\left\{2\ddot{a}
+\dot{a}^2\left[\frac{2f'(a)}{f(a)}+\frac{g'(a)}{g(a)}\right]+\frac{2f'(a)}{f(a)g(a)}\right\}.
\label{e13}
\end{equation}
The flare-out condition implies that the product $f(a)h(a)$ is an increasing function of $a$ so from Eq. (\ref{e11}) is immediate to see that the surface energy density is negative, which reflects the presence of exotic matter at the throat.
The tensions $\vartheta_{\theta }$, $\vartheta_{z}$ and the energy density $\sigma$ 
satisfy the equation
\begin{equation}
\vartheta_{\theta }-\vartheta_{z}=\frac{f'(a)h(a)-f(a)h'(a)}{f'(a)h(a)+f(a)h'(a)}\sigma.
\label{e14}
\end{equation}
Equations (\ref{e11}), (\ref{e12}) and (\ref{e13}) plus one of  the equations of state $\vartheta_{\theta }(\sigma)$ or $\vartheta_{z}(\sigma)$ determine the dynamics of the shell. We emphasize that these dynamic equations are valid under the assumption that the velocity of the throat is small so that the geometry outside the shell remains static and the emission of gravitational waves can be neglected.

\section{Stability analysis}\label{stability}

In this Section, we analyze the stability of static solutions under perturbations preserving the symmetry. The equations for a static shell are obtained by replacing  $\dot{a}=0$ and $\ddot{a}=0$ in Eqs. (\ref{e11}), (\ref{e12}) and (\ref{e13}):
\begin{equation}
\sigma = -\frac{1}{8\pi\sqrt{g(a)}}\left[\frac{f'(a)}{f(a)}+\frac{h'(a)}{h(a)}\right],
\label{e15}
\end{equation}
\begin{equation}
\vartheta_{\theta} = -\frac{1}{4\pi\sqrt{g(a)}}\frac{f'(a)}{f(a)},
\label{e16}
\end{equation}
\begin{equation}
\vartheta_{z} = -\frac{1}{8\pi\sqrt{g(a)}}\left[\frac{f'(a)}{f(a)}+\frac{h'(a)}{h(a)}\right].
\label{e17}
\end{equation}
Eqs. (\ref{e16}) and (\ref{e17}) can be recast in the form
\begin{equation}
\vartheta_{\theta}=\frac{2f'(a)h(a)}{f(a)h'(a)+f'(a)h(a)}\sigma,
\label{e18}
\end{equation}
\begin{equation}
\vartheta_{z}=\sigma ;
\label{e19}
\end{equation}
thus, for a given value of the throat radius, the functions $f$, $g$ and $h$ determine the equations of state $\vartheta_{\theta}(\sigma)$ and $\vartheta_{z}(\sigma)$ of the exotic matter on the shell. Because the analysis is restricted to small velocity perturbations around the static solution, the evolution of the exotic fluid in the shell can be considered as a succesion of static states. Thus we assume that the form of the equations of state for the static case is preserved in the dynamic case, thus $\vartheta_{\theta}(\sigma)$ and $\vartheta_{z}(\sigma)$ are given by Eqs. (\ref{e18}) and (\ref{e19}). Then, replacing Eqs. (\ref{e11}) and (\ref{e12}) in Eq. (\ref{e18}) (or Eqs. (\ref{e11}) and (\ref{e13}) in Eq. (\ref{e19})), the following second order differential equation for $a(\tau )$ is obtained:
\begin{equation} 
2g(a)\ddot{a}+g'(a)\dot{a}^{2}=0.
\label{e22}
\end{equation}
It is easy to see that 
\begin{equation} 
\dot{a}(\tau )\sqrt{g(a(\tau))}=\dot{a}(\tau _{0})\sqrt{g(a(\tau _{0}))},
\label{e23}
\end{equation}
satisfies Eq. (\ref{e22}), with $\tau _{0}$ an arbitrary (but fixed) time. Integrating both sides of Eq. (\ref{e23}) from $\tau _{0}$ to $\tau $ and making the substitution 
$da=\dot{a}(\tau )d\tau $, we have 
\begin{equation} 
\int^{a(\tau )}_{a(\tau _{0})}\sqrt{g(a)}da=\dot{a}(\tau _{0})
\sqrt{g(a(\tau _{0}))}(\tau - \tau _{0}).
\label{e25}
\end{equation}
Replacing Eq. (\ref{e1b}) in Eq. (\ref{e25}), and performing the integration, this gives
\begin{equation}
\tau -\tau_{0} = \frac{\mid\gamma\mid \sqrt{x_{s}}\left(\ln x_{s}\right)^{5/8}}
{2^{5/8}\dot{a}(\tau_{0})\sqrt{g(a(\tau_{0}))}}\\
\left\{\Gamma\left[\frac{3}{8},\left(1-\frac{\ln a(\tau)}{\ln x_{s}}\right)^{2}\frac{\ln x_{s}}{2}\right]-\Gamma\left[\frac{3}{8},\left(1-\frac{\ln a(\tau_{0})}{\ln x_{s}}\right)^{2}\frac{\ln x_{s}}{2}\right]\right\},
\label{e26} 
\end{equation}
where $\Gamma(\alpha,z)\equiv\int_{z}^{\infty}u^{\alpha-1}e^{-u}du$ is the incomplete gamma function. Inverting this relation the evolution of the radius of the throat $a(\tau )$ turns to be
\begin{equation}
a(\tau)=x_{s}\exp\left\{-\sqrt{2\Gamma^{-1}\left[\frac{3}{8},A+B(\tau-\tau_{0})
\right]\ln x_{s}}\right\},
\label{e27} 
\end{equation}
with $\Gamma(\omega)\equiv\int_{0}^{\infty}u^{\omega-1}e^{-u}du$ the Euler gamma function, $\Gamma^{-1}(\alpha,z)$ the inverse (for fixed $\alpha$) of the incomplete gamma function,
\begin{equation}
A=\Gamma \left[\frac{3}{8},\left(1-\frac{\ln a(\tau_{0})}
{\ln x_{s}}\right)^{2}\frac{\ln x_{s}}{2}\right],
\label{e27a} 
\end{equation}
and
\begin{equation}
B=\frac{2^{5/8}\dot{a}(\tau_{0})\sqrt{g(a(\tau_{0}))}}
{\mid\gamma\mid\sqrt{x_{s}}\left(\ln x_{s}\right)^{5/8}}.
\label{e27b} 
\end{equation}
Eq. (\ref{e27}) remains valid as long as the velocity given by Eq. (\ref{e23}) is small. The qualitative behaviour of $a(\tau)$ is not easily seen from Eq. (\ref{e27}), but it can be obtained from Eqs. (\ref{e22}) and (\ref{e23}). Following Eq. (\ref{e22}), the sign of the acceleration of the throat is determined by the sign of the derivative of the radial function $g(a)$. It is clear that $g'(a)$ is negative if $a\in(x_{1},x_{2})$, with
\begin{equation}
x_{1,2}=\sqrt{x_{s}}\exp \left(\pm\frac{1}{2}\sqrt{\ln x_{s}(\ln x_{s}-1)}\right),
\label{e28}
\end{equation}
where the minus and the plus signs correspond, respectively, to $x_{1}$ and $x_{2}$. As $\ln x_{s}\gg 1$, we have that $x_{1}\approx 1$ and $x_{2}\approx x_{s}$, so that the acceleration of the throat is always positive. From Eq. (\ref{e23}),  we can see that the sign of the initial velocity determines the sign of the velocity at any time. Therefore, when the initial velocity $\dot{a}(\tau_{0})$ is positive the wormhole throat presents an accelerated expansion and if $\dot{a}(\tau_{0})$ is negative it has a decelerated collapse. We want to emphasize that the temporal evolution allows only for these two possibilities because the sign of the velocity, which is given by its initial value, remains unchanged.

\section{Discussion}\label{discu}

In this paper we have studied thin-shell traversable Lorentzian wormholes constructed from the metric of a static global cosmic string with boost invariance. An observer outside the throat could not distinguish by local measurements between the wormhole geometry considered here and the global string spacetime. We have presented a general stability analysis under perturbations preserving the symmetry of these wormholes. We have assumed that the equations of state of the exotic matter at the throat are linear relations between the surface tensions and the surface energy density, and their form is the same in the dynamic case as in the static one. We have found that the temporal evolution of the wormhole throat is basically determined by the sign of  its initial velocity. If it is positive, the wormhole throat expands monotonically; when it is negative, the wormhole throat collapses to the core radius in finite time; and, if it is null, the wormhole throat remains at rest. Then, we conclude that there are no oscillating solutions. We have explicitly shown static solutions for any value of the radial coordinate but these solutions are not stable under radial perturbations in the velocity. We remark that this analysis is valid as long as the velocity of the throat is  much smaller than the velocity of light to guarantee a static spacetime outside the throat, so that  the emission of gravitational waves can be neglected. Global string geometries without boost invariance  \cite{senba} have a metric that can be written in a similar form of those studied in Ref. \cite{eisi}. It is straightforward to see that for thin-shell wormholes constructed from these strings the stability properties are the same as those obtained here and in Ref. \cite{eisi}. All these results seem to be a general  consequence of cylindrical configurations and the equations of state assumed. 

\section*{Acknowledgments}

This work has been supported by Universidad de Buenos Aires and CONICET. Some calculations were done with the help of the package GRTensorII (freely available at http://grtensor.org).

\end{document}